\documentclass[twocolumn,preprintnumber,amsmath,amssymb]{revtex4}
\usepackage{dcolumn}
\usepackage{bm}
\usepackage{graphicx}

\begin{document}
\title{Scattering of Scalar Waves by Schwarzschild Black Hole Immersed in Magnetic Field}
\author{ Juhua Chen } \email{jhchen@hunnu.edu.cn}
\author{ Hao Liao}
\author{ Yongjiu Wang }
\affiliation{College of Physics and Information Science,  Key
Laboratory of Low Dimensional Quantum Structures and Quantum Control
of Ministry of Education, Hunan Normal University, Changsha, Hunan
410081}

\begin{abstract}
The magnetic field is one of the most important constituents of the
cosmic space and one of the main sources of the dynamics of
interacting matter in the universe. The astronomical observations
imply the existence of a strong magnetic fields of up to
$10^4-10^8G$ near supermassive black holes in the active galactic
nuclei and even around stellar mass black holes. In this paper, with
the quantum scattering theory, we analysis the Schr\"{o}edinger-type
scalar wave equation of black hole immersed in magnetic field and
numerically investigate its absorption cross section and scattering
cross section. We find that the absorption cross sections oscillate
about the geometric optical value in the high frequency regime.
Furthermore in low frequency regime, the magnetic field makes the
absorption cross section weaker and this effect is more obviously on
lower frequency brand. On the other hand, for the effects of
scattering cross sections for the black hole immersed in magnetic
field, we find that the magnetic field makes the scattering flux
weaker  and its  width narrower in the forward direction. We find
that there also exists the glory phenomenon along the backforward
direction. At fixed frequency, the glory peak is higher and the
glory width becomes narrower due to the black hole immersed
in magnetic field. 
\end{abstract}

\maketitle

\section{Introduction}
It is well known that general relativity and quantum mechanics are
incompatible in their current form. However, after Hawking found
that black holes can emit, as well as scatter, absorb, and that the
evaporation rate is proportional to the total absorption cross
section.  A lot of scholars are interest in the absorption of
quantum fields by black hole since 1970s.
 By using numerical methods, Sanchez \cite{Sanchez1,Sanchez} found that
 the absorption cross section of massless scalar wave exhibits oscillation
  around the geometry-optical limit characteristic of diffraction
 patterns by  Schwarzschild black hole. Unruh \cite{Unruh} showed that
 the scattering cross section for the fermion is exactly 1/8 of that for the
 scalar wave in the low-energy limit. By numerically solving the single-particle Dirac equation in
Painlev\'{e}-Gullstrand coordinates, Chris Doran et al \cite{Doran}
studied the absorption of massive spin-half particle by a small
Schwarzschild black hole and they found oscillations around the
classical limit whose precise form depends on the particle mass.
Crispino et al \cite{Crispino1} have computed numerically the
absorption cross section of electromagnetic waves for arbitrary
frequencies and have found that its high-frequency behavior is very
similar to that for massless scalar field by Schwarzschild black
hole. In last several years, Oliveria et al \cite{Oliveira} extended
to study the absorption of planar in a draining bathtub, the
absorption cross section of sound waves with arbitrary frequencies
in the canonical acoustic hole spacetime \cite{Crispino} and
electromagnetic absorption cross section from Reissner-Nordstr\"{o}m
black holes \cite{Crispino2}. Recently, absorption cross section (or
gray body factors) has been of interest in the context of
higher-dimensional using standard field theory in curved spacetimes
\cite{Das,Das1,Crispino4} and effective string model \cite{Gubser}.

The magnetic field is one of the most important constituents of the
cosmic space and one of the main sources of the dynamics of
interacting matter in the universe. In addition some other theories
\cite{Tyulbashev, Zhang, Han} imply the existence of a  strong
magnetic fields of up to $10^4-10^8G$ near supermassive black holes
in the active galactic nuclei and even around stellar mass black
holes. In order to make estimations of possible influence of the
magnetic field on the supermassive black holes, we need the two
parameters at hand: the magnetic field parameter $B$ and the mass of
the black hole $M$. Interaction of a black hole and a magnetic field
can happen in a lot of physical situations: when an accretion disk
or other matter distribution around black hole is charged; when
taking into consideration galactic and intergalactic magnetic
fields, and, possibly, if mini-black holes are created in particle
collisions in the brane-world scenarios. So astrophysics have highly
interest to investigate the magnetic fields around black holes
\cite{Konoplya1}. A magnetic field is important  as a background
field testing  black hole geometry. A magnetic field near a black
hole leads to a number of processes, such as extraction of
rotational energy from a black hole, known as the Blandford-Znajek
effect \cite{Blandford}, negative absorption (masers) of electrons
\cite{Aliev}. At the classical level, the magnetic perturbation can
also be described by its damped characteristic modes, which called
the quasinormal modes (QNMs)
\cite{Kokkotas,Kokkotas1,Noller,Konoplya2} which could be observed
in experiments, and by the scattering properties, which are encoded
in the S-matrix of the perturbation. All of these effects are
usually called the "fingerprints" of a black hole. In recent few
years, we all know that quasinormal modes of black holes has gained
considerable attention because of their applications in string
theory through the AdS/CFT correspondence.

In this paper we mainly focus on the scalar scattering process of
black hole immersed in magnetic field and  how the interaction of
black hole and strong magnetic field  effects on scalar absorption
and scattering cross sections. The outline of this paper is as
follows: In Sec.II, we set up scalar field equation black holes
immersed in a magnetic field and analysis effective potential. In
the Sec. III and IV, we concentrate on the absorption and scattering
cross section of the scalar wave by black holes immersed in a
magnetic field. In the last section, a brief conclusion is given.

\section{Scalar Field Equation and Effective Potential}
 The Diaz and Ernst solution \cite{Ernst} describing the black holes immersed in a
magnetic field takes the follow form:
\begin{eqnarray} \label{metric}
ds^2&=&\Lambda^2[(1-\frac{2M}{r})dt^2+(1-\frac{2M}{r})^{-1}dr^2 \nonumber \\
\newline &-&r^2d\theta^2]-\frac{r^2sin^2\theta}{\Lambda^2}d\phi^{2},
\end{eqnarray}
where the external magnetic field is determined by the parameter $B$
\begin{eqnarray}
\Lambda=1+\frac{1}{4}B^2r^2sin^2\theta,
\end{eqnarray}
and the unit magnetic field measured in $Gs$ is
$B_{M}=1/M=2.4\times10^{19}\frac{M_{Sun}}{M}$.

The general perturbation equation for the massless scalar field
$\Psi$ in the curve spacetime is given by
 \begin{eqnarray}
 \frac{1}{\sqrt{-g}}\partial_\mu(\sqrt{-g}g^{\mu\nu}
 \partial_\nu)\Psi=0.\label{K-G}
 \end{eqnarray}

For very strong magnetic fields in centres of galaxies or in
colliders, corresponds to $M\ll M$ in our units, so that one can
safely neglect terms higher than $B^2$ in Eq.(\ref{K-G}). Indeed, in
the expansion of $\Lambda^4$ in powers of B, the next term after
that proportional to $ B^2r^2$, is $\sim B^4r^4$ and, thereby, is
very small in the region near the black hole. The term $ B^4r^4$ is
growing far from black hole, and, moreover the potential in the
asymptotically far region is diverging, what creates a kind of
confining by the magnetic field of the Ernst solution. This happens
because the non-decaying magnetic field is assumed to exist
everywhere in the universe. Therefore it is clear that in order to
estimate a real astrophysical situation, one needs to match the
Ernst solution with a Schwarzschild solution at some large r.
Fortunately we do not need to do this for the scattering problem:
the scattering properties of astrophysical interest is stipulated by
the behavior of the effective potential in some region near black
hole, while its behavior far from black hole is insignificant
\cite{Konoplya}. In this way we take into consideration only
dominant correction due-to magnetic field to the effective potential
of the Schwarzschild black hole. By neglecting  terms $B^4$ and
higher order terms and separating the angular variables, we reduce
the wave equation (\ref{K-G}) to the Schr\"{o}edinger wave equation
The Klein-Gordon equation can be written in the spacetime
(\ref{metric}) as
\begin{eqnarray}
 \frac{1}{1-\frac{2M}{r}}\frac{\partial^2\Psi}{\partial t^2}-\frac{1}{r^2}\frac{\partial}{\partial r}
 [(1-\frac{2M}{r}) r^2\frac{\partial\Psi}{\partial r}]+\frac{1}{r^2}\nabla^2\Psi=0.\label{eq2}
 \end{eqnarray}
 The positive-frequency solutions of Eq.(\ref{eq2}) take as follows
 \begin{eqnarray}
 \Psi_{\omega lm}=[\psi_{\omega l}(r)/r]Y_{lm}e^{-i\omega t},\label{eq3}
 \end{eqnarray}
 where $Y_{lm}$ are  scalar spherical harmonic functions and $l$ and $m$ are the corresponding angular momentum quantum numbers.
 In this case, the functions $\psi_{lm}(r)$ satisfy  the follow
  differential equation
 \begin{eqnarray}
 (1-\frac{2M}{r})\frac{d}{dr}[(1-\frac{2M}{r})\frac{d\psi_{\omega l}}{dr}]+[\omega^2-V^{(l)}_{eff}(r)]\psi_{\omega l}=0, \label{eq4}
 \end{eqnarray}
  where
\begin{eqnarray}
V^{(l)}_{eff}(r)&=&
(1-\frac{2M}{r})[\frac{l(l+1)}{r^2}+\frac{2M}{r^3}+4B^2m^2]
\label{Veff}.
 \end{eqnarray}

 \begin{figure}
\begin{center}
\includegraphics{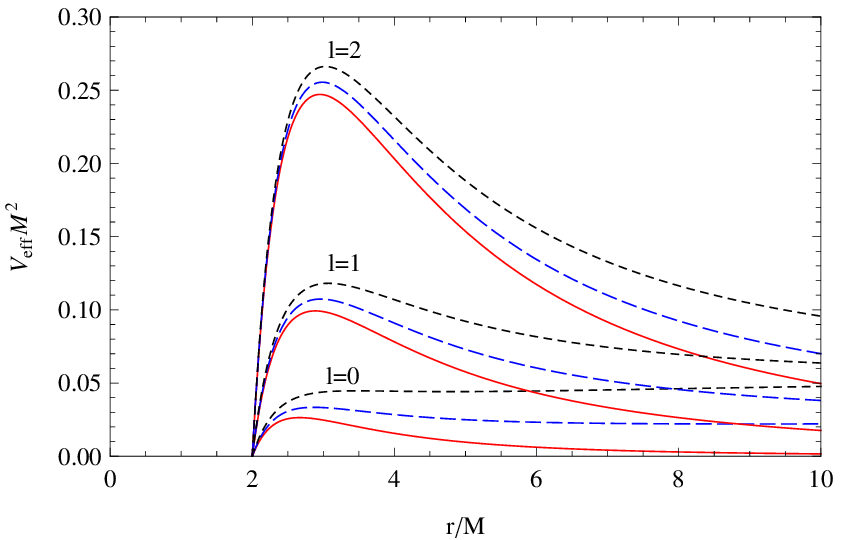}
\end{center}
\caption{(color online). The effective scattering  potential
$V_{eff}(r)$ given by Eq.\ref{Veff} for scalar waves by the black
hole immersed in magnetic field with $l=0,1,2$ for $B=0$ (red solid
line, i.e. Schwarzschild case) and $B=0.08$ (blue dashed line), and
$B= 0.12$ (black dotted line). }

\begin{center}
\includegraphics{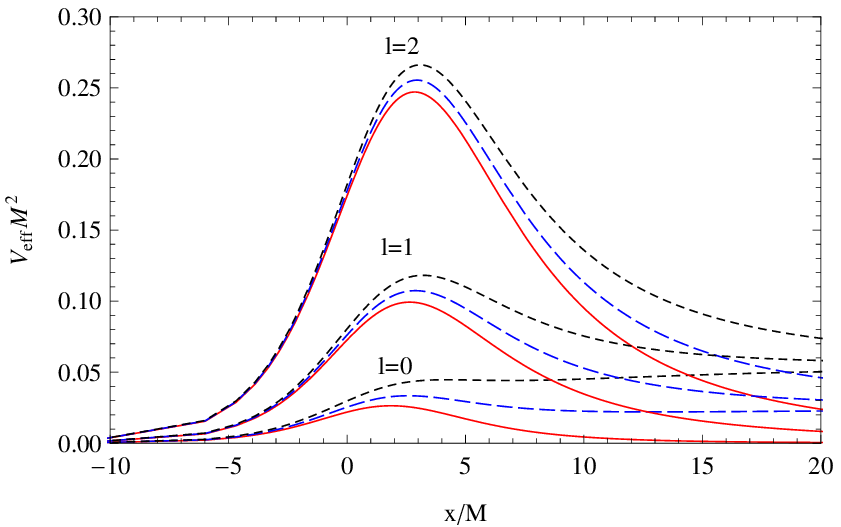}
\end{center}
\caption{(color online). The effective scattering  potential
$V_{eff}(x)$ given by Eq.\ref{Veff} for scalar waves by the black
hole immersed in magnetic field in tortoise coordinate with
$l=0,1,2$ for $B=0$ (red solid line, i.e. Schwarzschild case) and
$B=0.08$ (blue dashed line), and $B= 0.12$ (black dotted line). From
this figure, we can see the effective scattering potential
$V_{eff}(x)$ act as the typical scattering barrier in quantum
mechanics theory. }
\end{figure}

The effective potential $V^{(l)}_{eff}(r)$ is plotted in Fig.1 for
$l=0, 1, 2$. From this figure, we can see that the effective
potential $V^{(l)}_{eff}(r)$ depends only on the values of $r$,
angular quantum number $l$, ADM mass $M$, magnetic field $B$,
respectively, and that the peak value of potential barrier gets
upper and the location of the peak point ($r=r_p$) moves along the
right when the angular momentum $l$ increases. We  can find that the
the height of the effective scattering potential increases as the
angular momentum $l$ increases. If we introduce the tortoise
coordinate
\begin{eqnarray}
x= \int{(1-\frac{2M}{r})^{-1}dr},
\end{eqnarray}
The effective potentials $V^{(l)}_{eff}(r)$ are changed into
$V^{(l)}_{eff}(x)$, which are showed in Fig.2 for $l=0, 1, 2$, it's
obvious that they act as the typical scattering barriers in quantum
mechanics theory. We see that the peak value of potential barrier
gets upper and the location of the peak point ($x=x_p$) moves along
the right when the angular momentum $l$ increases. We  also find
that the height of the effective scattering barrier increases as the
magnetic field $B$ increases, at the same time we can see that the
height of the effective scattering barrier, affecting by the
magnetic field,becomes higher than that of Schwarzschild black hole.

After introducing this coordinate transition, we can obtain the
following Schr\"{o}dinger-type equation
\begin{eqnarray}
\frac{d^2\psi_{\omega
l}}{dx^2}+[\omega^2-V^{(l)}_{eff}(x)]\psi_{\omega l}=0.\label{eq5}
\end{eqnarray}

The perturbation must be purely ingoing at the black hole event
horizon $r=r_+$. So  while $ r\rightarrow r_+$ i.e. $x\rightarrow
-\infty$, we impose the boundary condition
\begin{eqnarray}
\psi_{\omega l}=A^{tr}_{\omega l}e^{-i\omega x}, & for &
 x\rightarrow -\infty.\label{b1}
\end{eqnarray}
 It is straightforward to check that in the original coordinate system (\ref{metric}) the ingoing solution $e^{-i\omega x}$ is well defined at
 $r=r_+$, whereas the out going solution $e^{+i\omega x}$ is divergent. Towards
 spatial infinity, the asymptotic form of the solution is
\begin{eqnarray}
\psi_{\omega l}&=&\omega x[A^{in}_{\omega
l}(-i)^{l+1}h^{(1)\ast}_{l}(\omega x) + A^{out}_{\omega
l}(i)^{l+1}h^{(1)}_{l}(\omega x)]\nonumber \\
&for& x\rightarrow +\infty,\label{b2}
\end{eqnarray}
where $h^{(1)}_{l}(\omega x)$ are  spherical Bessel functions of the
third kind \cite{Abrammowitz}, at the same time $A_{in}$ and
$A_{out}$ are complex constants. We note that $h^{(1)}_{l}(\omega
x)\approx (-i)^{l+1}e^{ix}/x$ as $x\rightarrow\infty$ and that the
effective potential goes to zero as  $x\rightarrow -\infty$, so we
obtain
\begin{eqnarray}
\psi_{\omega l}\approx \bigg\{
         \begin{array}{rrrr}
  A^{tr}_{\omega l}e^{-i\omega x},& for& x\rightarrow -\infty; \\
  A^{in}_{\omega l}e^{-i\omega x}+ A^{out}_{\omega l}e^{+i\omega x}, & for &  x\rightarrow
+\infty.
          \end{array}
\end{eqnarray}
with the conserved relation
\begin{eqnarray}
|A^{tr}_{\omega l}|^2+|A^{out}_{\omega l}|^2=|A^{in}_{\omega l}|^2
\end{eqnarray}

The phase shift $\delta _l$ is defined by
\begin{eqnarray}
e^{2i \delta _l}=(-1)^{l+1}A_{out}/A_{in}.\label{delt}
\end{eqnarray}
In order investigate the absorption cross section and scattering
cross section, we must numerically solve the radial equation
(\ref{eq5}) under the boundary conditions Eq.(\ref{b1}) and
Eq.(\ref{b2}), then compute the ingoing and outgoing coefficients
$A^{in}_{\omega l}$ and $A^{out}_{\omega l}$ by matching onto
Eq.(\ref{delt}) to give out the numerical phase shift.

\section{Absorption cross section}
 Base on the quantum mechanics theory, we know that the total absorption cross
section is
\begin{eqnarray}
\sigma_{abs}=\frac{\pi}{\omega^2}\sum_{l=0}^{\infty} (2l+1)(1-|e^{2i
\delta _l}|^2),\label{abs}
\end{eqnarray}
so we can define the partial absorption cross section as
\begin{eqnarray}
\sigma^{(l)}_{abs}=\frac{\pi}{\omega^2}(2l+1)(1-|e^{2i \delta
_l}|^2),\label{absp}
\end{eqnarray}
and  the  absorption cross section have relation
\begin{eqnarray}
\sigma_{abs}(\omega)=\sum_{l=0}^{\infty}\sigma^{(l)}_{abs}(\omega)=\frac{\pi}{\omega^{2}}\sum_{l=0}^{\infty}(2l+1)|T_{\omega
l}|^2.\label{T}
\end{eqnarray}
By using {\sl mathematica} program, we straightforwardly compute
values of ingoing and outgoing coefficients $A^{in}_{\omega l}$ and
$A^{out}_{\omega l}$. Then from Eq.(\ref{delt}), Eq.(\ref{abs})and
Eq.(\ref{absp}), we can simulate the partial absorption cross
sections and their total absorption cross sections of the scalar
field from the black hole immersed in magnetic field.

In Fig.3 we show the partial absorption cross sections
$\sigma_{abs}^{(l)}$, i.e. $l=0,1,2$, by the  black hole immersed in
magnetic field for different magnetic parameters $B=0.1$ and
$\Lambda=0,0.08$ and $0.12$. We find that the S-wave $(l=0)$
contribution is responsible for the nonvanishing cross section in
the zero-energy limit. Furthermore, by comparing different $l$
partial absorption cross section curves, we find that the larger the
value of $l$ is, the smaller the corresponding value of
$\sigma_{abs}^{(l)}$ is. This is compatible with the fact that the
scattering barrier $V_{eff}$ is bigger or larger values of $l$,
which is showed in Fig.1 and Fig.2. These properties are similar to
other black hole scattering system\cite{Sanchez,Crispino,Crispino3}.
On the other hand with phase-integral method, Andersson
\cite{Andersson} had gotten very similar results (see Fig.7
therein).

In order to consider effects of magnetic field on the partial
absorption cross section. In Fig.4 we plot the partial absorption
cross section for $l=0,1,2$ with $B=0$ ( i.e. Schwarzschild black
hole case), $B=0.08$  and $B= 0.12$. We see that the magnetic field
make the absorption weaker, even for low frequency mode. This is
agree with the fact that the magnetic field is stronger, the higher
value of the effective scattering  barrier peak is for a fixed value
of $l$, which can be seen in Fig.1 and Fig.2. But for high enough
values of the frequency, the magnetic field does not effect the
partial absorption cross section obviously. From Eq.(\ref{T}), we
know the absorption cross section have relation with the
transmission coefficients \cite{Decanini1}. This feature can be
tested the transmission coefficients in Fig.5, where we find that
high enough values of the frequency all transmission coefficients
with fixed $l$ tend to the unity. These properties help us
understand the absorption process better.

\begin{figure}
\begin{center}
\includegraphics{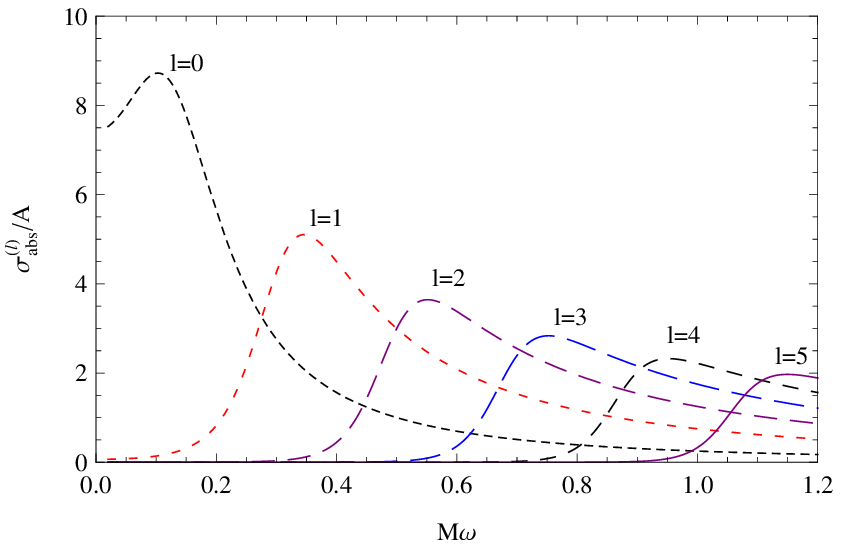}
\end{center}
\caption{(color online). The behavior of the partial absorption
cross section $\sigma_{abs}^{(l)}$, from $l=0$ to $l=5$ for scalar
waves by the black hole immersed in magnetic field.}

\begin{center}
\includegraphics{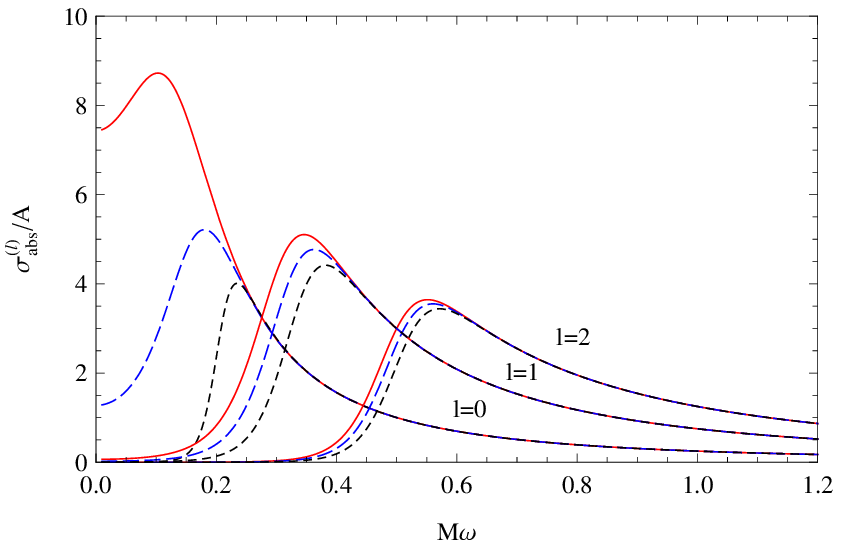}
\end{center}
\caption{(color online). The behavior of the partial absorption
cross section $\sigma_{abs}^{(l)}$, from $l=0,1,2$ for scalar waves
by the black hole immersed in magnetic field with $B=0$ (red solid
line, i.e. Schwarzschild black hole case) and $B=0.08$ (blue dashed
line), and $B= 0.12$ (black dotted line).}

\begin{center}
\includegraphics{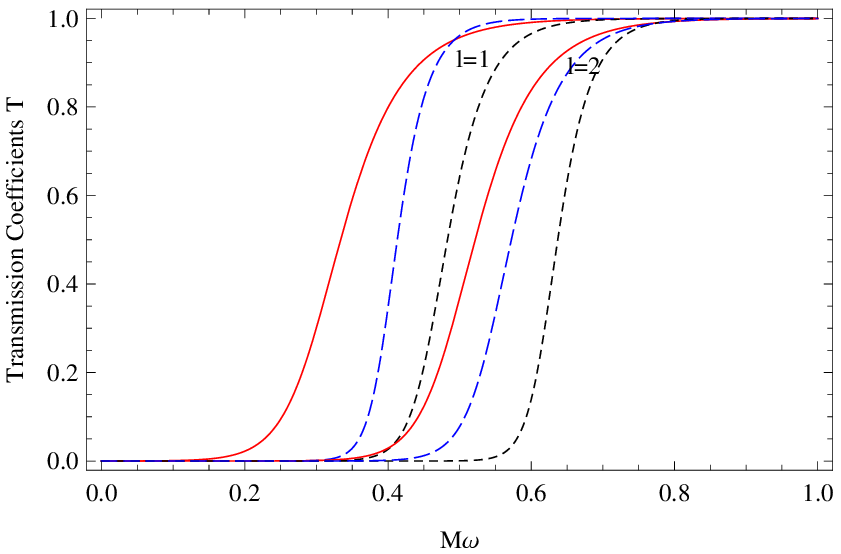}
\end{center}
\caption{(color online). The transmission coefficients with
$l=1,l=2$ are showed for different magnetic field  with $B=0$ (red
solid line, i.e. Schwarzschild black hole case) and $B=0.08$ (blue
dashed line), and $B= 0.12$ (black dotted line). }
\end{figure}

\begin{figure*}
\begin{center}
\includegraphics{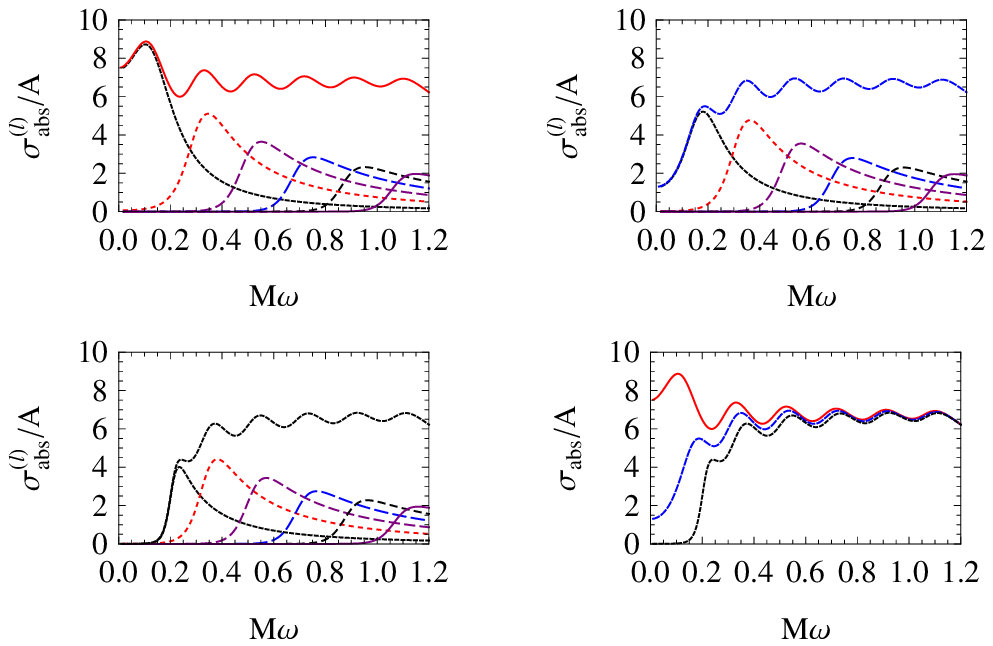}
\end{center}
\caption{(color online). The behavior of the partial absorption
cross section $\sigma_{abs}^{(l)}$  and $\sigma_{abs}^{total}$  by
the black hole immersed in magnetic field with $B=0$ (top-left i.e.
Schwarzschild black hole case), $B=0.08$ (top-right), $B= 0.12$
(bottom-left), and their corresponding total absorption cross
sections $\sigma_{abs}^{total}$ (bottom-right).}
\begin{center}
\includegraphics{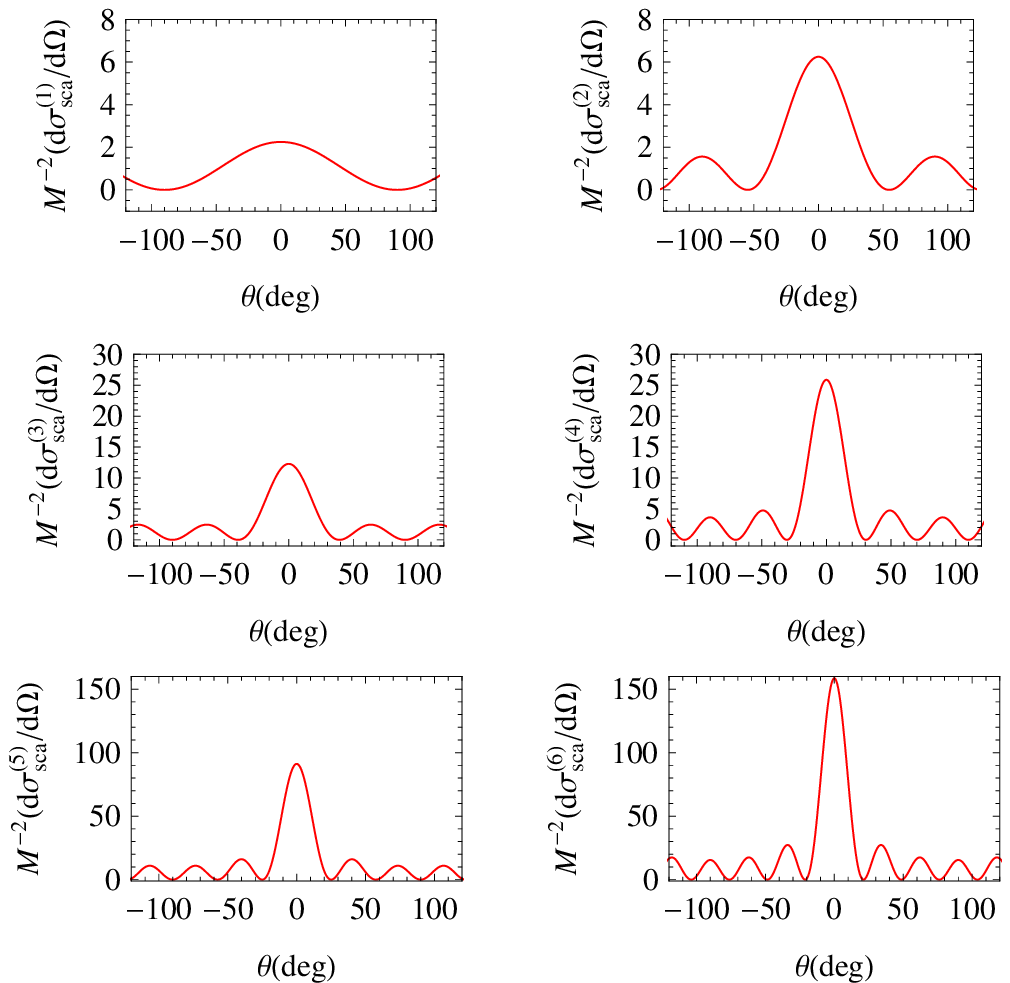}
\end{center}
\caption{(color online). The behavior of the partial scattering
cross sections $\sigma^{(l)}_{sca}$, from $l=1$ to $l=6$, at
$M\omega=1$ for the scalar wave is scattered by the black hole
immersed in magnetic field with $B=0.2$.}
\end{figure*}

In Fig.6 we plot total absorption cross sections $\sigma_{abs}$
which contribute from $l=0$ to $l=5$ by the black hole immersed in
magnetic field with fixed parameters $B=0$ ( i.e. Schwarzschild
case), $B=0.08$ and $B=0.12$. We can see that I) between the
intermediate regime $\omega M\sim (0.4,1)$, the contributions from
the partial absorption sections create a regular oscillatory
pattern. Each maximum in the oscillation of the total absorption
cross section is linked to the maximum of a particular partial wave.
II) If the wavelength of the incoming wave is much smaller than the
black hole horizon (i.e. $\omega M>>1$), the absorption cross
section tends to the geometry-optical limit of
$\sigma_{abs}^{hf}=\pi b^2_{c}$. This is verified by the total
absorption cross section for the massless scalar field which was
computed by Sanchez \cite{Sanchez} in last century. At the same
time, these properties are also found for electromagnetic wave
absorption cross section \cite{Crispino1} and for Fermion absorption
cross section in the Schwarzschild black hole \cite{Doran}.

In bottom-left position of Fig.6, we plot total absorption cross
sections for different values of magnetic parameters. We also
consider the contributions of the angular momentum from $l=0$ to
$l=5$ in Eq.(\ref{absp}). We can see that big values of the magnetic
parameter $B$ correspond to low total absorption cross section which
is consistent with the fact of the partial section in Fig.4. and the
scattering barrier which is showed in Fig.1 and 2. But we can find
that the absorption cross sections oscillate about the geometric
optical value in the high frequency regime. However in low frequency
regime, the magnetic field makes the absorption cross section
weaker, i.e. the magnetic makes obvious effect on lower frequency
brand, not on high frequency brand. We note that this is a general
result for massless scalar waves in Reissner-Nordstr\"{o}m black
hole \cite{Crispino3} and for the minimally-coupled massless scalar
wave in stationary black hole spacetimes \cite{Higuchi}. There are
similar properties for total absorption section from the charged
black hole coupling to Born-Infeld electrodynamics \cite{chen} and
dark energy \cite{Liao} .

\section{Scattering cross section}
From the quantum mechanics theory, it's well known that the
scattering amplitude is expressed as
\begin{eqnarray}
f(\theta)=\frac{1}{2i\omega}\sum_{l=0}^{\infty} (2l+1)[e^{2i \delta
_l}-1]P_{l}(cos\theta).
\end{eqnarray}
From this scattering amplitude, we can give the differential
scattering cross section immediately
\begin{eqnarray}
\frac{d\sigma}{d\Omega}=|f(\theta)|^2.
\end{eqnarray}
At last we can define the scattering and  absorption cross sections
\cite{Gotfried,Dolan}
\begin{eqnarray}
\sigma_{sca}&=& \int \frac{d\sigma}{d\Omega}
d\Omega=\frac{\pi}{\omega^2}\sum_{l=0}^{\infty} (2l+1)|e^{2i \delta
_l}-1|^2,\label{sca}
\end{eqnarray}
so the partial scattering cross section is
\begin{eqnarray}
\sigma^{(l)}_{sca}=\frac{\pi}{\omega^2}(2l+1)|e^{2i \delta
_l}-1|^2.\label{scap}
\end{eqnarray}

In order to simulate the scattering cross sections
(\ref{sca})-(\ref{scap}), we must numerically solve differential
equation (\ref{eq5}) under boundary conditions (\ref{b1}) and
(\ref{b2}), to obtain numerical values for the phase shifts via
Eq.(\ref{delt}).

Figures 7 show the partial scattering cross section a function of
angle for six different partial waves from $l=1$ to $l=6$. By
comparing these figures, we can see that, when the $L$ increases,
the flux is preferentially scattered in the forward direction, i.e.
the scattering angle width become narrower. At the time  a more
complicated pattern arises and we find a damping oscillation
pattern. The similar properties are observed for black hole
scattering \cite{Dolan,Futterman,Dolan1}. The explanation for the
physical origin of the oscillations can be found in
Ref.\cite{Matzner}.

Figures 8, 9 and 10 compare the scattering cross sections for the
Scharzschild black hole with the black hole immersed in magnetic
field with $B=0.2$ and $0.3$. We find that the magnetic field makes
the scattering flux weaker  and its  width narrower in the forward
direction. In the other words, the scalar field scattering becomes
more diffusing due to the black hole immersed in magnetic field. In
Fig.10 we can see that there exists the glory phenomenon along the
backforward direction \cite{Dolan,Crispino3}. At fixed frequency,
the glory peak is higher and the glory width becomes narrower due to
the black hole immersed in magnetic field. So we can find that even
the scalar field scattering becomes more diffusing due to the black
hole immersed in magnetic field, but the glory phenomenon along the
backforward direction becomes better for astronomy observation.

\begin{figure}
\begin{center}
\includegraphics{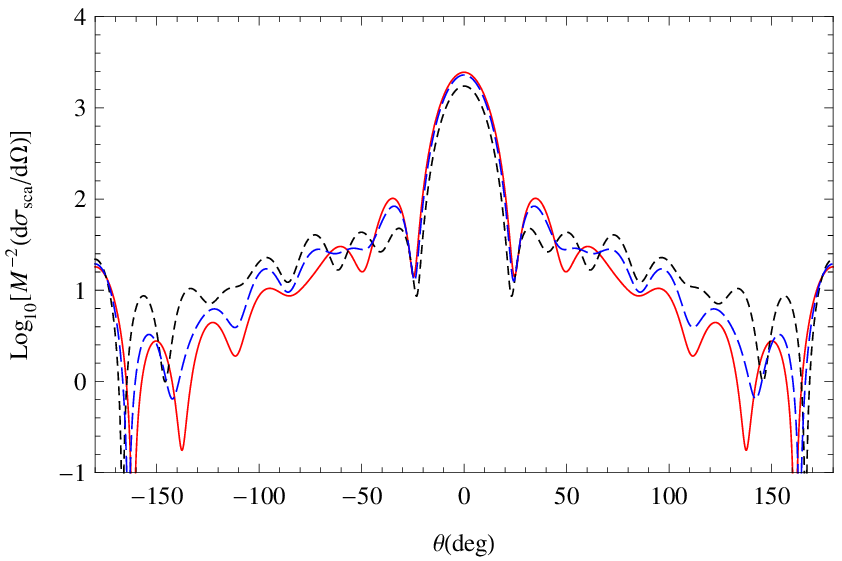}
\end{center}
\caption{(color online). The behavior of the total scattering  cross
sections $\sigma^{(l)}_{sca}$ at $M\omega=1$ between
($-180^{\circ}$-$180^{\circ}$) for the scalar wave is scattered by
the black hole immersed in magnetic field with $B=0$ (red solid
line, i.e. Schwarzschild case) and $B=0.2$ (blue dashed line), and
$B= 0.3$ (black dotted line).}

\begin{center}
\includegraphics{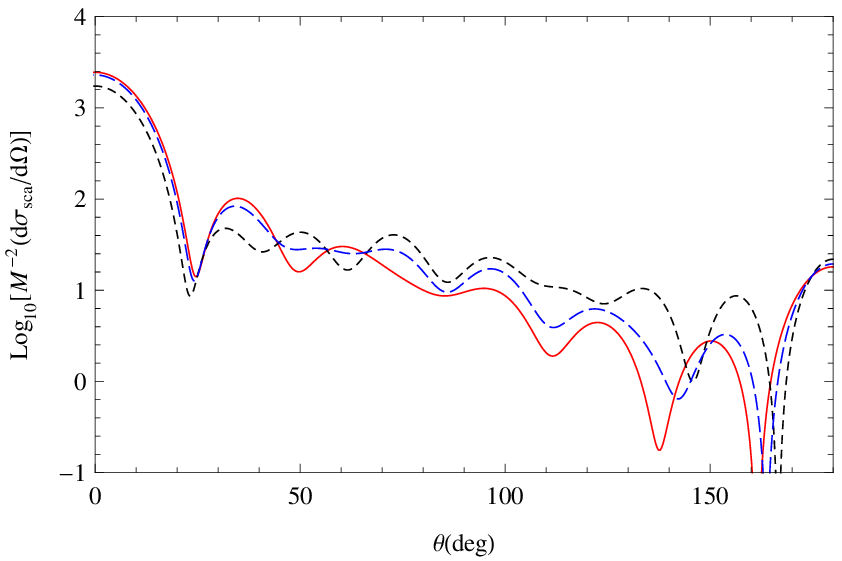}
\end{center}
\caption{(color online). The behavior of the total scattering  cross
sections $\sigma^{(l)}_{sca}$ at $M\omega=1$ between
($0^{\circ}$-$180^{\circ}$)  for the scalar wave is scattered by the
black hole immersed in magnetic field with $B=0$ (red solid line,
i.e. Schwarzschild case) and $B=0.2$ (blue dashed line), and $B=
0.3$ (black dotted line).}

\begin{center}
\includegraphics{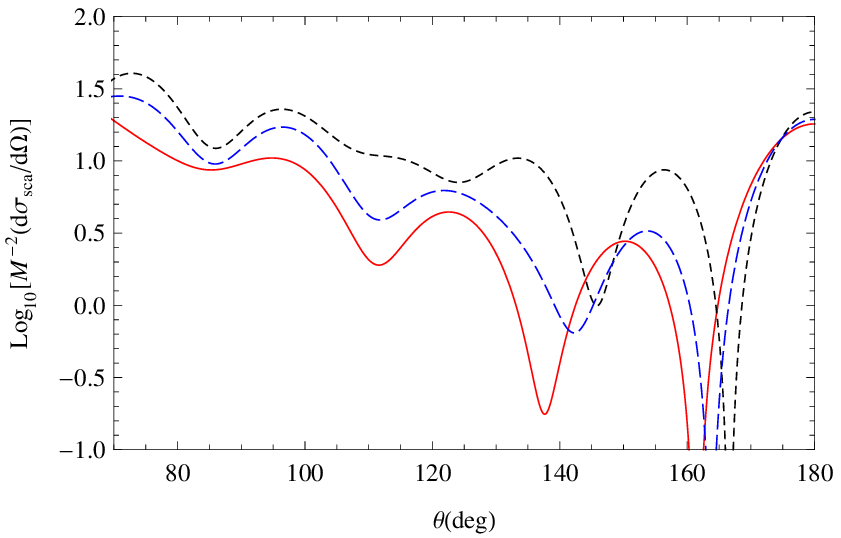}
\end{center}
\caption{(color online). The behavior of the total scattering  cross
sections $\sigma^{(l)}_{sca}$ at $M\omega=1$ between
($60^{\circ}$-$180^{\circ}$) for the scalar wave is scattered by the
black hole immersed in magnetic field with $B=0$ (red solid line,
i.e. Schwarzschild case) and $B=0.08$ (blue dashed line), and $B=
0.12$ (black dotted line).}
\end{figure}

\section{Conclusions}
In this paper we have investigated the scattering and absorption
cross section of the scalar wave by the black hole immersed in
magnetic field.  We found that the magnetic parameter $B$ makes the
absorption cross section lower which is consistent with the fact of
the scattering barrier which is showed in Fig.1 and 2.  We also
found that the absorption cross sections oscillate about the
geometric optical value in the high frequency regime. However in low
frequency regime, the magnetic field makes the absorption cross
section weaker and this effect is more obviously on lower frequency
brand. For the effects of the scattering cross sections for the
black hole immersed in magnetic field, we found that the magnetic
field makes the scattering flux weaker and its scattering width
narrower in the forward direction. At the same time  we found that
there also exists the glory phenomenon along the backforward
direction. At fixed frequency, the glory peak is higher and the
glory width becomes narrower due to the black hole immersed in
magnetic field. So  the glory phenomenon along the backforward
direction becomes better for astronomy observation.

Just as the Brazil physicist Crispino et al \cite{Crispino3} have
pointed out: " In principle, highly accurate measurements of, for
example, the gravitational wave flux scattered by a black hole could
one day be used to estimate the black hole¡¯s charge. A more
immediate possibility is that scattering and absorption patterns may
be observed with black hole analog systems created in the
laboratory. Even if experimental verification is not forthcoming, we
hope that studies of wave scattering by black holes will continue to
improve our understanding of how black holes interact with their
environments."
\section{Acknowledgments}
This project is supported by the National Natural Science Foundation
of China under Grant No.10873004, the State Key Development Program
for Basic Research Program of China under Grant No.2010CB832803 and
the Program for Changjiang Scholars and Innovative Research Team in
University, No. IRT0964.

\end{document}